# Prototype of Front-end Electronics for PandaX-4ton Experiment

Shuwen Wang, Zhongtao Shen, Keqing Zhao, Changqing Feng, and Shubin Liu

*Abstract*—At the China Jinping Underground Laboratory, the Particle AND Astrophysical Xenon phase IV (PandaX-4ton) in planning is a dark matter direct detection experiment with dual-phase xenon detector as an upgrade of the second phase of the experiment, PandaX-II. In this paper, the prototype of the front-end electronics of PandaX-4ton is presented. The front-end electronics consist of the high-gain preamplifier cards and the eight-channel digitizers with 14-bit resolution and 1 GSps sampling rate for waveform digitization. The clock synchronization circuit within the digitizer is well-designed to align all the PMT channels. The digitizer also contains gigabit fiber to exchange data with trigger and data acquisition system. The specification of effective number of bits f the digitizer is about 9.7 b at 148 MHz, and the integral nonlinearity of the digitizer ranges from -4 least significant bit (LSB) to +4 LSB, and the differential nonlinearity ranges from -0.6 LSB to +0.6 LSB. The performance of the front-end electronics can meet the requirements for the PandaX-4ton.

## I. INTRODUCTION

THE Particle AND Astrophysical Xenon phase IV (PandaX-4ton) in planning at China Jinping Underground Laboratory (CJPL) [1] in Sichuan is a dark matter direct detection experiment with dual-phase xenon [2] as an upgrade of the second phase of the experiment, PandaX-II [3]. PandaX-4ton is currently leading in the search for incoming weakly interacting massive particles (WIMPs). As shown in Fig.1, when WIMPs collide with the dual-phase xenon, photons (S1 signal) are produced in the liquid xenon. Ionized electrons are then drifted vertically upward by an induced drift field and extracted into the gas by an extraction field, producing the electroluminescence (S2 signal). The S1 signal and S2 signal are collected by the photomultiplier tubes (PMTs) on the top and bottom of the detector [4], and then fed into front-end electronics.

PandaX-4ton as the upgrade of PandaX-II has 512 channels and lower threshold. It requires front-end electronics to have a high-level integration. The rising edge of S1 signal is about 2ns-3ns. So to precisely obtain the S1 signal wave information, we have designed a multi-channel digitizer with 14-bit resolution and 1 GSps sampling rate.

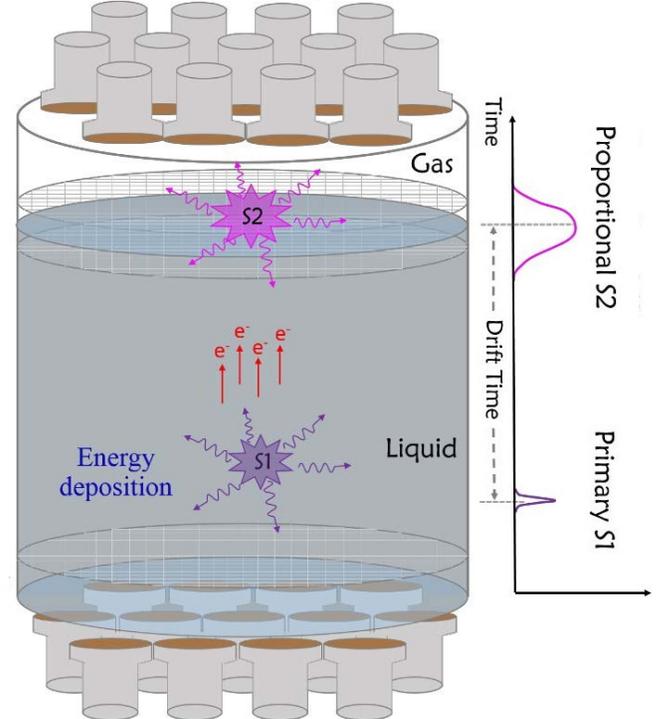

Fig. 1. Dark matter detection in dual-phase xenon detector.

## II. FRONT-END ELECTRONICS SYSTEM

As illustrated in Fig.2, front-end electronics is composed of preamplifier cards and front-end digital module (FDM).

The preamplifier card is mainly for amplifying the signals from PMTs before the signals are fed into the FDM.

Each FDM is integrated with four 14-bit 1GSps Analog-to-Digital Converter (ADCs), which can digitize up to 8 detector channels. The Field Programmable Gate Array (FPGA) on the FDM controls the alignment of data from ADCs and processes the aligned data. FPGA also communicates with trigger and DAQ electronics through optical fiber. It can receive trigger signals and then transmits valid data to trigger and DAQ electronics. Besides a spare local clock source, FDM also receives the reference clock which is recovered through optical fiber for synchronization purpose.

This work was supported by the Fundamental Research Funds for the Central Universities (Grant WK2030040097). (Corresponding author: Zhongtao Shen)

S. Wang, Z. Shen, K. Zhao, C. Feng, and S. Liu are with State Key Laboratory of Particle Detection and Electronics, and Department of Modern Physics, University of Science and Technology of China, Hefei, 230026, China (e-mail: henzt@ustc.edu.cn).

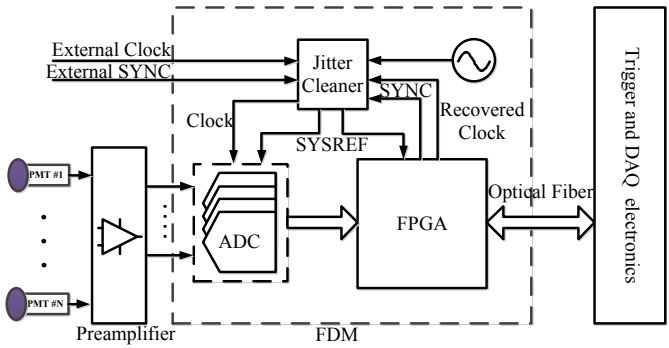

Fig. 2. Structure of front end electronics.

*A. Preamplifier Circuit*

The preamplifier card is shown in Fig.3. The preamplifier circuit is cascaded by two stage amplifiers. The first stage can amplify the signal 2 times, and then signal is amplified 10 times by the second stage to match the maximum range of ADC.

-3dB bandwidth of the preamplifier circuit is about 250 MHz as shown in Fig.4, which matches the analog bandwidth of the S1 signal.

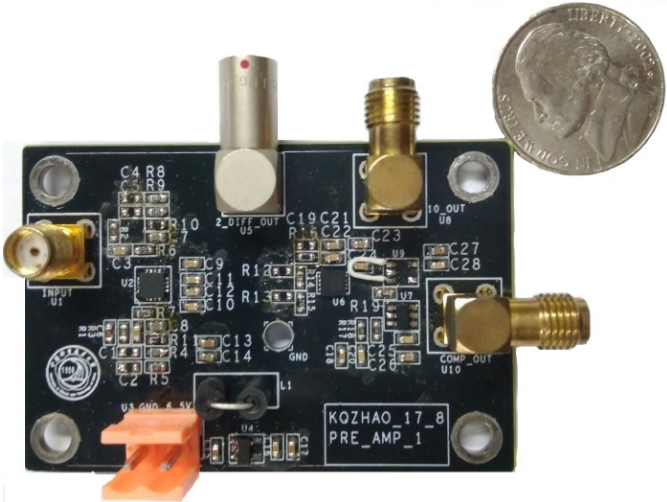

Fig. 3. Photograph of the preamplifier card.

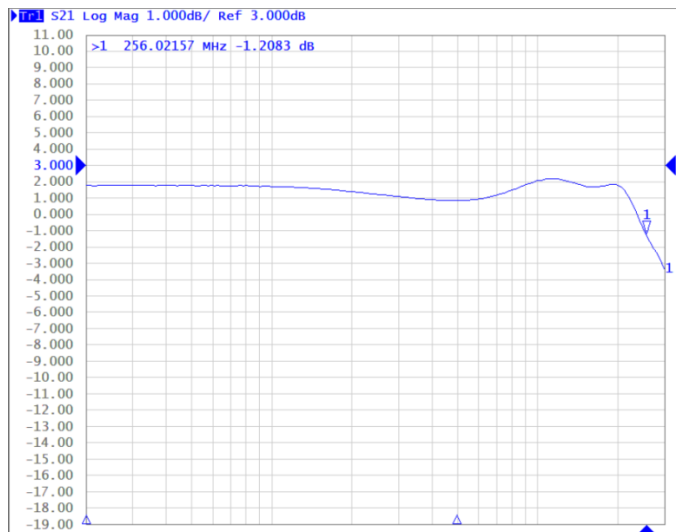

Fig. 4. Test result of the preamplifier card.

*B. FDM*

A 6U PXI-based FDM prototype has been implemented as illustrated in Fig.5. Each FDM is integrated with four 14-bit 1GSps JESD204B [5] compatible ADCs.

The JESD204B serialized interface for the high-speed ADCs supports the high bandwidth, high speed and multi-channel applications, while greatly reducing the number of digital IOs needed and thus easing board layout. The total bandwidth of the JESD204B interface can also be separated into multiple channels based on the requirements of the application again without requiring additional pins. The JESD204B system also has mechanism to achieve deterministic latency and synchronization across the serial link. Using the JESD204B compatible ADCs improves the integration of the front-end electronics system.

Each ADC has two digitization channels. To digitize up to all the 512 PMT channels, there will be 64 FDMs.

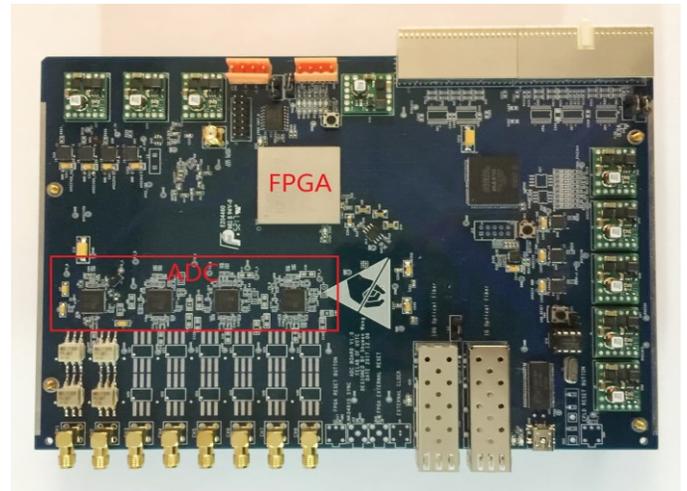

Fig. 5. Test result of the preamplifier card.

*C. Synchronization mechanism*

The synchronization mechanism of the front-end electronics system is shown in Fig.6. All FDMs work with a system synchronous clock of 125MHz. The synchronous clock is from the trigger and DAQ electronics through the optical fiber, and then recovered by FPGA clock and data recovery (CDR) module. Then the recovered clock passes through a JESD204B compliant clock jitter cleaner to retain the accurate sampling clock and generate JESD204B SYSREF [5] signal while simultaneously suppressing the higher offset frequency phase noise. The phase alignment of the sampling clocks at each ADC is critical for maintaining alignment of the sampling instances. And the SYSREF signal is the most important signal for achieving repeatable system latencies and synchronization in the JESD204B compatible systems. In this way, FDM can synchronize all the ADCs within the same FDM.

FDMs also receive SYNC signal from the trigger and DAQ electronics. SYNC signal makes all FDMs reset at the same time. This ensures that all the channel dividers of the different jitter cleaners start with the same edge of the clock. Doing that, all the output clocks are edge aligned, and all channels are synchronized.

FDM can also receive external clock and SYNC signal. It can be used as debugging.

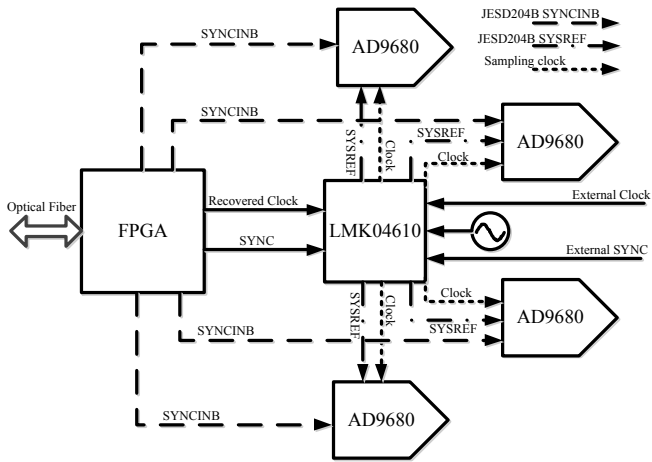

Fig. 6. Schematic of multi-device synchronization.

## III. TEST

The static performance and the dynamic performance were tested to verify the design of FDM. Testing environment is shown in Fig. 7.

### A. Static Performance Tests

A sine wave signal with frequency of 9.7MHz was used for the ADC's static performance test. As shown in Figs. 8 and Figs. 9, the integral nonlinearity (INL) of the ADC ranges from -4 least significant bit (LSB) to +4 LSB, and the differential nonlinearity (DNL) ranges from -0.6 LSB to +0.6 LSB.

### B. Dynamic Performance Tests

Sine wave signals with frequency ranging from 30.5 MHz to 148 MHz were used for the dynamic performance test. The results of the ENOB are show in Fig. 10. The ENOB is 10.4 b at 30.5 MHz and degrades to about 9.7 b at 148 MHz.

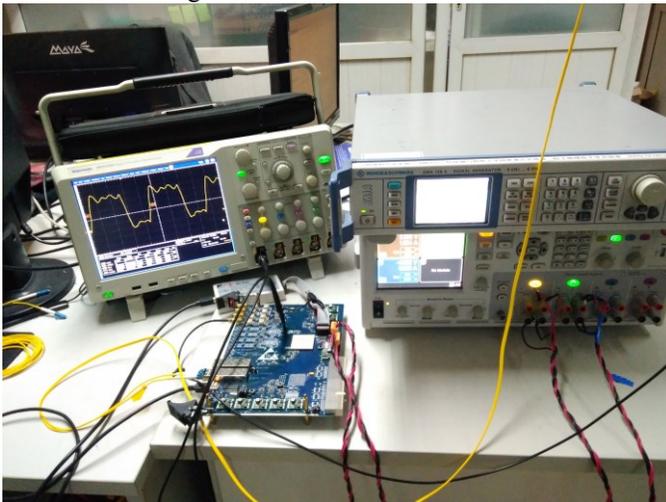

Fig. 7. Test result of the preamplifier card.

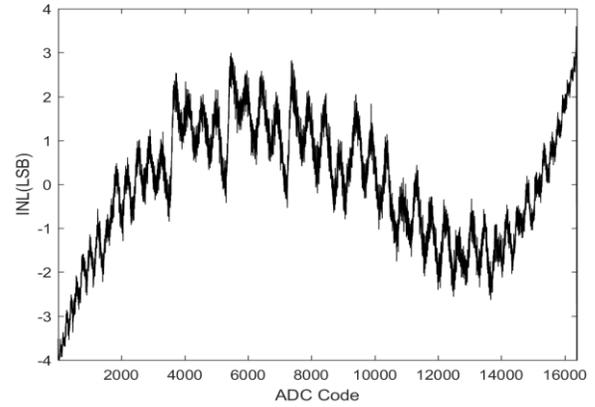

Fig. 8. Test results of INL.

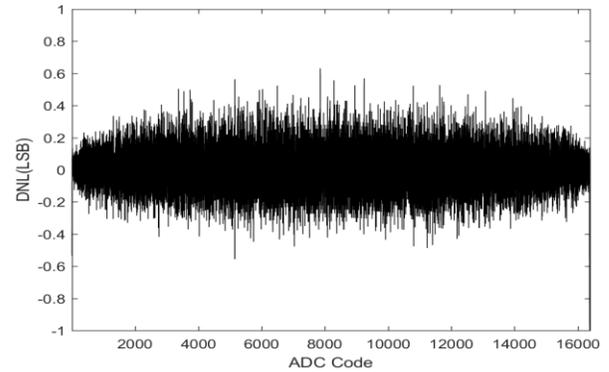

Fig. 9. Test results of DNL.

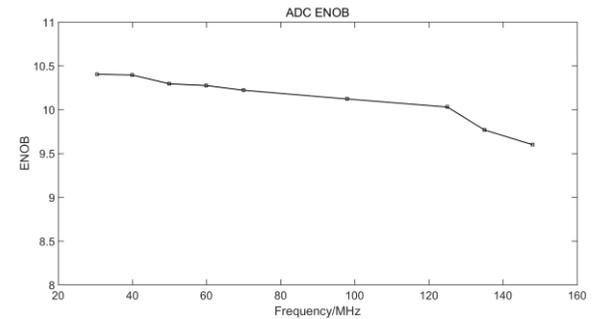

Fig. 10. Test results of ENOB.

## IV. CONCLUSION

Prototype of front-end electronics for PandaX-4ton experiment is proposed in this paper. It is composed of preamplifier cards and FDMs. Each FDM can digitize up to 8 detector channels at 1 GSps sampling rate with 14-bit resolution. Compared to PandaX-II experiment, it has advantages of simplicity, high resolution, high sampling rate. Test results show that the performance of the FDM can meet the requirements for PandaX-4ton experiment.

## REFERENCES

[1] Y. Wu et al, "Measurement of cosmic ray flux in the China JinPing underground laboratory," Chin. Phys. C, vol. 37, no. 8, 086001 (2013).